\documentclass[english,aps,pra,twocolumn,letterpaper]{revtex4-1}

\usepackage[T1]{fontenc}

\usepackage{graphicx}
\usepackage{epstopdf}
\usepackage{amssymb}
\usepackage{bm}
\usepackage{amsmath}
\usepackage{color}
\usepackage{rotating}
\usepackage{subfigure}

\usepackage[english]{babel}
\usepackage{blindtext,tikz}
\usetikzlibrary{calc}

\newcommand{\be}{\begin{equation}}
\newcommand{\ee}{\end{equation}}

\newcommand{\ketbra}[2]{\left|#1\right>\!\left<#2\right|}
\newcommand{\abs}[1]{\left|#1\right|}
\newcommand{\kk}{\kappa}
\newcommand{\yy}{\gamma}

\newcommand{\ww}{\omega}
\newcommand{\eps}{\epsilon}
\newcommand{\la}{\lambda}

\begin{document}

\title{Enhanced qubit readout using locally-generated squeezing and inbuilt Purcell-decay suppression}

\author{Luke C. G. Govia}
%\email[Electronic address: ]{govial@physics.mcgill.ca}
\address{Department of Physics, McGill University, Montreal, Quebec, Canada.}
\author{Aashish A. Clerk}
\address{Department of Physics, McGill University, Montreal, Quebec, Canada.}

\begin{abstract}
We introduce and analyze a dispersive qubit readout scheme where two-mode squeezing is generated directly in the measurement cavities.  The resulting suppression of noise enables fast, high- fidelity readout of naturally weakly coupled qubits, and the possibility to protect strongly coupled qubits from decoherence by weakening their coupling.  Unlike other approaches exploiting squeezing, our setup avoids the difficult task of transporting and injecting with high fidelity an externally-generated squeezed state.  Our setup is also surprisingly robust against unwanted non-QND backaction effects, as interference naturally suppresses Purcell decay:  the system acts as its own Purcell filter.  Our setup is compatible with the experimental state-of-the-art in circuit QED systems, but the basic idea could also be realized in other systems.
\end{abstract}

\maketitle

\section{Introduction}

Quantum states of radiation that have quadrature noise suppressed below the vacuum level, the so called squeezed states \cite{QuantumNoise,Walls:1983aa}, have a long history in quantum optics and quantum information science. Their key utility has been in interferometric measurements, where they can be used to dramatically improve precision \cite{Giovannetti1330}.  This has far ranging applications from gravitational wave detection \cite{Caves:1980aa,Braginsky:1992kq,Aasi2013}, to optomechanics \cite{Iwasawa:2013aa,Peano:2015aa}, and even biology \cite{Taylor:2013aa}.
More recently, squeezed states have been studied as a means to enhance qubit readout in quantum computing \cite{Barzanjeh2014,Didier2015}.
The focus has been on superconducting quantum circuits, where readout is typically performed using a dispersive coupling to a microwave cavity
\cite{Chow:2014fk,Hatridge:2013zr,Jeffrey:2014zr,Murch:2013rt,Riste2013,Roch:2014jk,Steffen2013}, and where significant squeezing of microwave photons can be readily produced \cite{Eichler:2011aa,Zhong:2013aa,Eichler:2014aa,Murch:2013aa,Toyli:aa}.
In principle, by injecting squeezed radiation into a measurement cavity, one can significantly enhance readout fidelity
\cite{Barzanjeh2014,Didier2015,Didier:2015aa}. In practice however it is extremely difficult to transport and inject squeezed states into a measurement cavity without losses:  state of the art transfer efficiencies are typically 70\% at best \cite{Murch:2013aa,Toyli:aa}.  As a result, the practical utility of injected squeezing for
enhanced qubit measurement is severely limited.

In this work, we analyze a realistic setup that overcomes this omnipresent limitation.  We introduce a two-cavity setup for dispersive qubit readout
where two-mode squeezed radiation is generated \emph{in situ}.  We thus completely circumvent the difficult task of faithfully transporting a squeezed state to the system being measured.  Combined with a dynamical symmetry of the system \cite{Didier2015}, this setup leads to a parametrically enhanced signal-to-noise ratio (SNR) and greatly reduced measurement times.  Our scheme is particularly effective in the case of a weak coupling between the readout cavity and the qubit:  here, we find that our scheme fundamentally changes the scaling of the SNR with coupling strength, such that it becomes independent of coupling strength at long times.  As a result, one can extract information about a weakly coupled qubit at the same rate as a conventional setup having a strong, optimally-tuned qubit-cavity coupling.  This remarkable feature of our setup is not possible if one simply injects squeezing into a traditional dispersive-readout cavity.

%%%%%%%%%%%%%%
\begin{figure}[t]
	\centering
\subfigure{
\includegraphics[height=0.5\columnwidth]{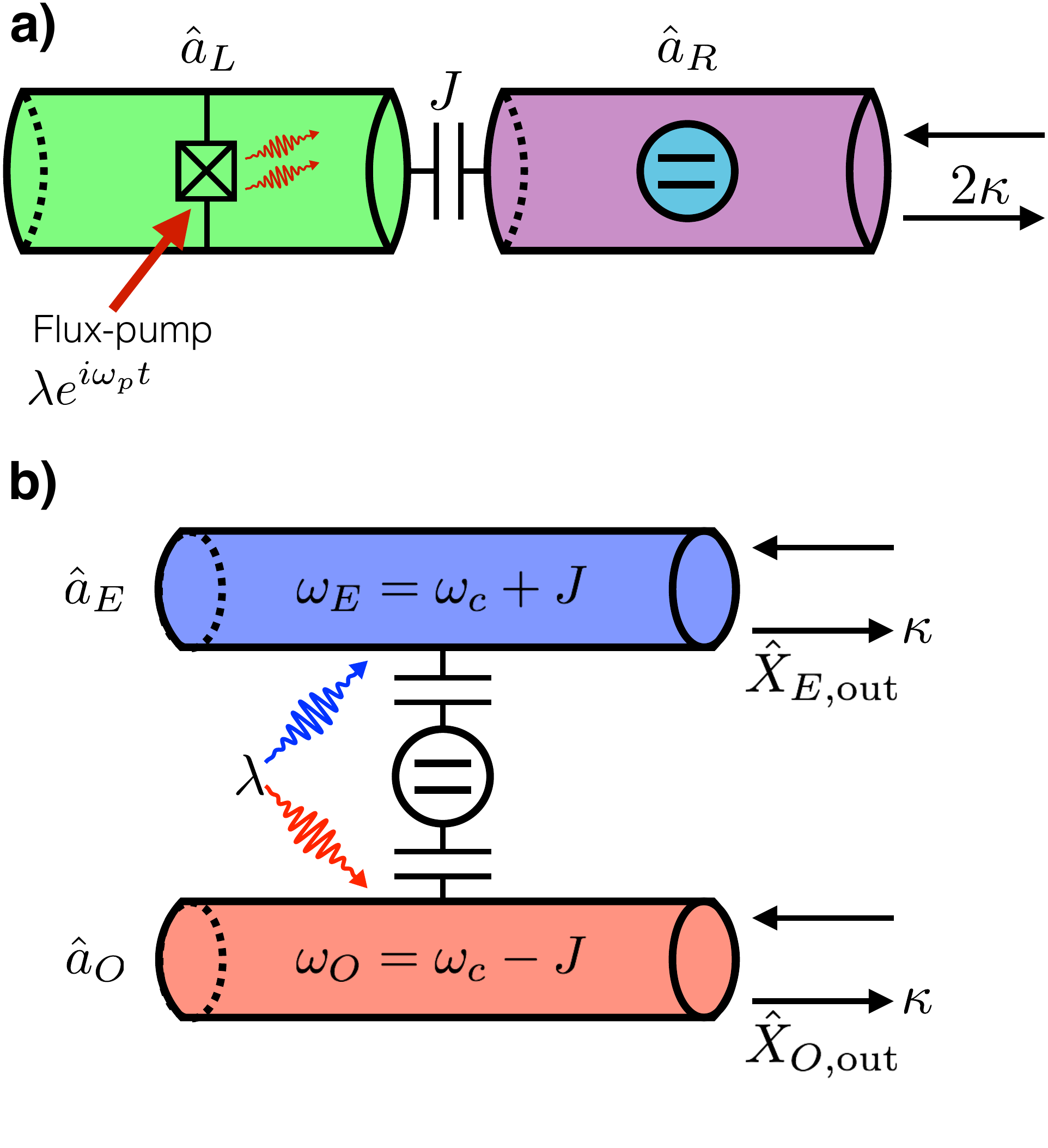}}
\subfigure{
\includegraphics[height=0.5\columnwidth]{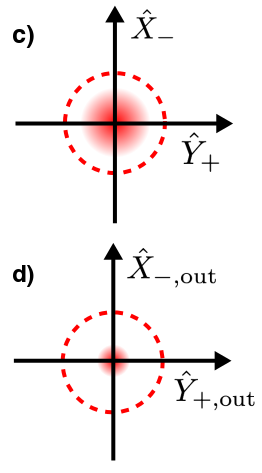}}
\caption{{\bf a)} The ISTMS setup, consisting of two tunnel-coupled cavities, one of which is coupled to a qubit and the input-output port, and the other flux-pumped by an imbedded SQUID. {\bf b)} Equivalently, the system can be thought of in terms of its eigenmodes, and the flux-pumping results in a two-mode squeezing interaction between them. {\bf c)} intracavity and {\bf d)} output state within the squeezed squeezed subspace defined by joint quadratures of the cavity normal modes (c.f. Eq.~(\ref{eqn:Ham2})), in the absence of any coupling to the qubit. The plots show that when compared to the vacuum state (dashed circle) the intracavity state is never squeezed by more than a factor of two, while the output state can be arbitrarily squeezed.}
\label{fig:setup}
\end{figure}
%%%%%%%%%%%%%%%%%%

The remarkable enhancement of measurement in weakly coupled systems is not just of academic interest.
Weakly coupled qubit-cavity systems are an emerging topic of current research, with recent developments in the coupling of a single-electron spin to a microwave cavity \cite{Petersson2012,Viennot2015,Stockklauser2015} having projected best-case coupling strengths that
are much weaker than typical cavity dissipation rates, putting them firmly in the weak coupling regime.  In addition,
new circuit QED designs have been proposed that leverage weak coupling to suppress photon shot noise dephasing \cite{Zhang2016}.
Our in-situ two-mode squeezing setup (ISTMS) would allow an exponential enhancement of measurement rates in such systems, in a way that would
be impossible even if one were able to inject external squeezing perfectly.

Our analysis reveals that our ISTMS setup possesses other highly desirable features.  Clearly, one potential drawback of generating squeezing in the measurement cavity is the possibility that the complementary amplified quadrature fluctuations in the cavity will cause extraneous backaction on the qubit.  Remarkably, such effects are greatly suppressed in our system, as is standard Purcell decay of the qubit due to simple cavity vacuum fluctuations.  This is because of a subtle interference effect, which causes the two-cavity system to act like a built-in Purcell filter.

%%%%%%%%%%%%%%%%
\section{ISTMS Setup}

We start with a setup that is somewhat analogous to the successful Bose-Hubbard dimer amplifier \cite{Eichler:2014aa}: two tunnel coupled cavities, denoted left and right, where the left cavity has an embedded SQUID, and the right cavity is coupled to an input-output transmission line. To this we add a qubit coupled to the right cavity via a Jaynes-Cummings interaction, and flux-pump the SQUID embedded in the left cavity \cite{Yamamoto:2008aa}. The ISTMS setup is shown in Fig.~\ref{fig:setup}a).

The Hamiltonian describing this system is ($\hbar = 1$ from here on)
\begin{align}
\hat{H} = \hat{H}_{\rm res} + \hat{H}_{\rm paramp}+ \hat{H}_q + \hat{H}_{\rm JC}
\end{align}
where $\hat{H}_{\rm res} = \ww_c \hat{a}_L^\dagger\hat{a}_L + \ww_c \hat{a}_R^\dagger\hat{a}_R + J\left(\hat{a}_L^\dagger\hat{a}_R+\hat{a}_R^\dagger\hat{a}_L\right)$ describes the tunnel coupled cavities (coupling strength $J$), with $\hat{a}_{L/R}$ ($\hat{a}_{L/R}^\dagger$) the lowering (raising) operators for the left/right cavities. $\hat{H}_q = \ww_q\hat{\sigma}_z/2$ is the qubit self-Hamiltonian, and $ \hat{H}_{\rm JC} = g\left(\hat{a}_R^\dagger\hat{\sigma}^- + \hat{\sigma}^+\hat{a}_R\right)$ is the Jaynes-Cummings coupling between the qubit and the right cavity. Here $\hat{\sigma}_z$ is the Pauli-Z matrix describing the qubit energy eigenstates, and $\hat{\sigma}^{\pm}$ are the raising and lowering operators for the qubit.

Finally, $\hat{H}_{\rm paramp} = i\la e^{i\ww_pt}\hat{a}_L^2 + h.c.$ is the parametric drive on the left cavity (amplitude $\la$) induced by flux-pumping the embedded SQUID at a frequency $\ww_p$ \cite{Yamamoto:2008aa}. In the simplest setup, higher-order cavity nonlinearities will also be induced by the flux-pumped SQUID for large enough drive amplitude $\la$. However, in contemporary experiments the effective nonlinearity is reduced, either by introducing an additional linear inductance in the circuit, or by using a SQUID array, such that even up to strong drive amplitudes only the parametric term is relevant \cite{Eichler:2014aa,Eichler2014}.

As we focus on the regime where  $J \gg g , \lambda$
the tunnel-coupled cavities are best described by their eigenmodes, $\hat{a}_E = (\hat{a}_L+\hat{a}_R)/\sqrt{2}$, $\hat{a}_O = (\hat{a}_R-\hat{a}_L)/\sqrt{2}$, which we refer to as the even and odd modes (see Fig.~\ref{fig:setup}(b)).  The corresponding normal mode frequencies are
$\ww_{E/O} = \ww_c \pm J$ . The flux pumping of the left cavity results in a two-mode squeezing interaction between these eigenmodes, which is resonant for $\ww_p = 2\ww_c$.

We move into an interaction picture, such that each cavity normal mode frequency is shifted to zero.
Setting $\ww_q = \ww_c$, we then make a standard dispersive transformation
to eliminate the qubit-cavity interaction to leading order; here, $J$ plays the role of a
large cavity-qubit detuning.  This results in the Hamiltonian
\begin{align}
\hat{H}_{D} = \chi\left(\hat{a}_E^\dagger\hat{a}_E - \hat{a}_O^\dagger\hat{a}_O\right)\hat{\sigma}_z - i\la\left(\hat{a}_E\hat{a}_O - \hat{a}^\dagger_E\hat{a}^\dagger_O\right), \label{eqn:Ham}
\end{align}
where $\chi = g^2/(2J)$ is the dispersive coupling, and where we have neglected terms that are higher-order in powers of $g / J, \lambda / J$.  We have also used the condition  $\la/J \ll 1$ to allow a rotating wave approximation where non-resonant terms coming from the parametric drive are dropped (see \ref{sec:AppTM} and \ref{sec:ApDis} for more details).

Eq.~(\ref{eqn:Ham}) shows clearly that the qubit is dispersively coupled to both cavity normal modes, with dispersive couplings that only differ in sign.
As discussed in Ref.~\cite{Didier2015}, this situation is highly advantageous if one wants to exploit squeezing, as it gives rise to a dynamical symmetry that has been termed a ``quantum mechanics free subspace'' \cite{Tsang:2012aa}; we discuss this more below.  Unlike Ref.~\cite{Didier2015}, this symmetry is achieved without having to explicitly engineer two distinct qubit-cavity couplings, giving us a distinct practical advantage.

The coupling between the right cavity to its readout transmission line is described using standard input-output theory \cite{QuantumNoise}, with an energy decay rate $2\kk$. Hence, the decay rate for each of the normal modes is $\kk$. Focusing on the case $J \gg \kk$, the $\hat{a}_E$ and $\hat{a}_O$ modes are well resolved in frequency, and thus each couple to an independent set of transmission line modes (see \ref{sec:inout}).

To understand how our system can be used for dispersive qubit readout we define $\hat{a}_{\pm} = (\hat{a}_E\pm\hat{a}_O)/\sqrt{2}$ (in our interaction picture), and rewrite Eq.~(\ref{eqn:Ham}) in terms of the quadrature operators $\hat{X}_{\pm}= (\hat{a}_{\pm}+\hat{a}_{\pm}^\dagger)/\sqrt{2}$ and $\hat{Y}_{\pm}= -i(\hat{a}_{\pm}-\hat{a}_{\pm}^\dagger)/\sqrt{2}$
\begin{align}
\hat{H}_{D} = \chi\left(\hat{X}_+\hat{X}_- + \hat{Y}_+\hat{Y}_-\right)\hat{\sigma}_z + \la\left(\hat{X}_+\hat{Y}_+-\hat{X}_-\hat{Y}_-\right). \label{eqn:Ham2}
\end{align}
The first term in this Hamiltonian describes a quantum-mechanics-free subsystem interaction \cite{Didier2015,Tsang:2012aa} that mediates a qubit dependent rotation in the effective phase-space plane defined by $\{\hat{X}_-,\hat{Y}_+\}$. A coherent driving of the  $\hat{Y}_{+}$ quadrature will be rotated by the interaction with the qubit, resulting in a non-zero value of $\langle \hat{X}_{-} \rangle$ that is qubit state dependent.  This can be detected via a homodyne measurement of the output field leaving the cavity, thus measuring the state of the qubit.

The second term in Eq.~(\ref{eqn:Ham2}) describes the generation of in-situ squeezing that is unique and so crucial to our setup.  It effectively
damps {\it both} the  $\{\hat{X}_-,\hat{Y}_+\}$ quadratures without adding noise, thereby generating intracavity two-mode squeezing.  While the intracavity squeezing remains weak, the squeezing of the {\it output} field quadratures can be much stronger, due to the well-known interference between the noise leaving the cavity and the vacuum noise that is promptly reflected at the cavity entrance \cite{QuantumNoise}, as is shown in Fig.~\ref{fig:setup}c) and d). The net result is that the vacuum noise is strongly reduced in any linear combination of $\{\hat{X}_-,\hat{Y}_+\}$ quadratures, and in particular the combination that has maximal information on the qubit state.  Unlike schemes using single-mode squeezing and dispersive interactions, one never has to worry about the anti-squeezed quadrature corrupting the measurement results.

\begin{figure}
\includegraphics[width=0.8\columnwidth]{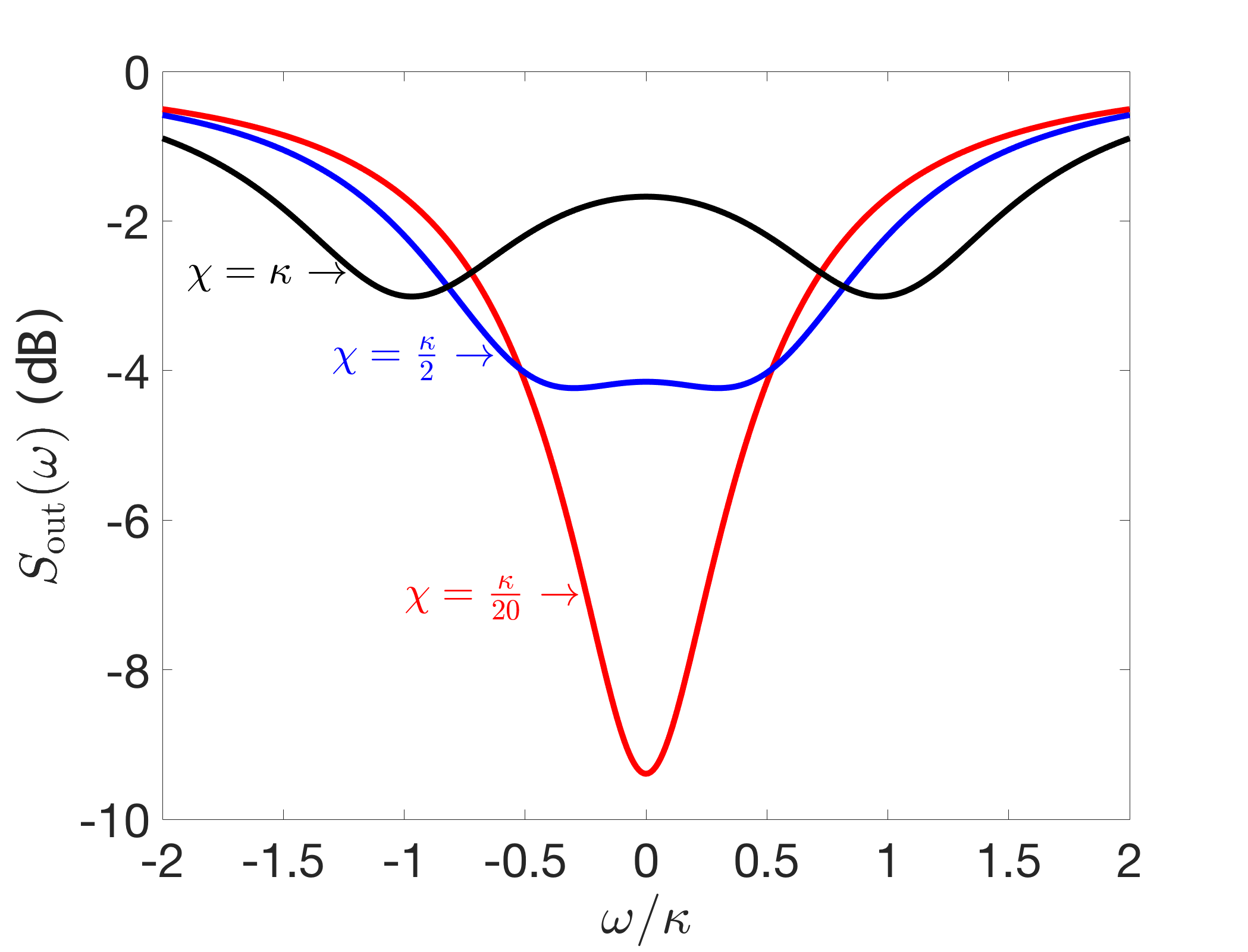}
\caption{Output squeezing spectrum of the joint quadrature $\hat{X}_{-,{\rm out}} = \left(\hat{X}_{E,{\rm out}} + \hat{X}_{O,{\rm out}}\right)/\sqrt{2}$ for the ISTMS setup (see \ref{sec:inout}), with $\la = \kk/4$. As $\chi \rightarrow 0$ the spectrum approaches a single Lorentzian, while for stronger coupling it is the sum of two Lorentzians centered at $\pm \chi$. Even for strong coupling the noise is always squeezed below zero point, as the qubit back-action does not mix amplified noise into the output.}
\label{fig:SqzSpec}
\end{figure}

While for $\chi = 0$, Eq.~(\ref{eqn:Ham2}) can produce arbitrarily large amounts of output two mode squeezing as one approaches the parametric instability at $\la = \kk/2$, for non-zero $\chi$, the interaction with the qubit will disrupt the needed interference.  The result is a double-peaked squeezing spectrum, with Lorentzian peaks at $\omega = \pm\chi$ (see Fig.~\ref{fig:SqzSpec}). This effect is negligible for weak coupling, and even at stronger couplings the output noise remains squeezed below zero point.  We stress that the interaction with the qubit never causes a mixing of squeezed and anti-squeezed quadratures (as occurs if one simply injects single-mode squeezing into a dispersively-coupled readout cavity \cite{Barzanjeh2014,Didier2015}).

%%%%%
%%%

%%%%%%%%%%%%%%%%%%%%%%%%%
\section{Measurement rate in the long time limit}

We first consider the measurement properties of our system in the limit where one integrates the homodyne current associated with the cavity
output field for a time $\tau \gg 1 / \kappa$.  In this limit, the SNR generically behaves as $\sqrt{\Gamma \tau}$, where $\Gamma$ is the measurement rate (i.e.~the rate at which information is obtained on the qubit state) \cite{Clerk:2010aa}.  For a standard dispersive readout (i.e.~a single cavity and a simple coherent state drive), one makes a homodyne measurement of the output phase quadrature.  The measurement rate satisfies
\begin{align}
	\Gamma_{\rm SM} =
\Bigg\{\begin{array}{ll}
		 4 \bar{n} \kk, &  {\rm if }\ \chi = \kappa/2 \\
		 2 \bar{n} \left(\chi/\kk\right)^2\kk, & {\rm if }\ \chi \ll  \kappa/2.
	\end{array}
	\label{eqn:SNRSM}
\end{align}
Here $\bar{n}$ is the average intracavity photon number induced by the coherent driving of the cavity.
The first line represents the optimal case, where one has a strong, optimally tuned cavity-qubit coupling.  The second line is the result for
a weakly coupled qubit. One cannot increase $\bar{n}$ without bound to increase $\Gamma_{\rm SM}$ for weak coupling, as $\bar{n}$ is bounded by the critical photon number \cite{Blais2004}, above which detrimental back-action on the qubit occurs. Hence, new ideas are needed to overcome the slow measurement rates associated with weakly-coupled qubits.

We now turn to our ISTMS setup. The measurement again involves coherently driving the cavity, and integrating the appropriate output homodyne current for a time $\tau$.  The coherent drive should displace the cavity $\hat{Y}_{E/O}$ quadratures, while the output homodyne measurement corresponds to monitoring the $\hat{X}_{-}$ collective quadrature.  The parametric drive $\lambda$ is kept on during the measurement to generate the desired squeezing (taking $\la < \kk/2$ to ensure the system stays stable).  Further details on how the collective quadrature is measured (using standard techniques) are provided in \ref{sec:inout}.

For long integration times, we find that the measurement rate is:
\begin{align}
	\Gamma_{\rm ISTMS}  =  \frac{8 \bar{n}_0\chi^2}{\chi^2+\left(\frac{\kk}{2} - \la\right)^2}\kk, \label{eqn:SNR}
\end{align}
Here, $\bar{n}_0$ is the intracavity photon number associated with the displacement induced by the coherent drive.
%It is given by
%\begin{align}
%n_0 = \frac{\kk\abs{\beta}^2}{\left(\la+\frac{\kk}{2}\right)^2+\chi^2},
%\label{eqn:n0}
%\end{align}
%with $\sqrt{2}\abs{\beta} = \left<\hat{Y}_{+,{\rm in}}\right>$ the amplitude of the coherent input photon flux. Note that $n_0$ decreases as $\abs{\la}$ increases, such that it is smaller for ISTMS readout than for single-mode readout.
The total average intracavity photon number $\bar{n} \equiv \langle \hat{a}_E^\dagger \hat{a}_E + \hat{a}_O^\dagger \hat{a}_O \rangle =
\bar{n}_0 + \bar{n}_{\rm sqz}$ also includes a contribution $\bar{n}_{\rm sqz}$ from the squeezing.  This is given by
$\bar{n}_{\rm sqz} = \la^2 / \left[ \left(\kk/2\right)^2-\la^2 \right]$.
%\begin{align}
%	\bar{n}_{\rm sqz} = \la^2 / \left[ \left(\kk/2\right)^2-\la^2 \right].
%%	 \xrightarrow{\chi \ll \kk} \frac{\kk}{4\chi} - 1.
%\end{align}

\section{Weak coupling enhancement}

For weak couplings $\chi \ll \kappa$, the qubit-induced disruption of squeezing (i.e.~$\chi^2$ term in denominator of Eq.~(\ref{eqn:SNR})) is weak, and the ISTMS scheme enables dramatic enhancements of the measurement rate.  To see this, we first note that once the parametric drive is increased past  $\la \sim \kk/2 - \chi$, $\Gamma_{\rm ISTMS}$ becomes independent of $\la$, and saturates to a value that is independent of the dispersive coupling $\chi$.  For the threshold value of $\la = \kk/2 - \chi$ we have
\begin{align}
	\Gamma_{\rm ISTMS} = 4 \bar{n}_0\kk .
	\label{eq:SNRWeak}
\end{align}
Strikingly, this expression is independent of the dispersive coupling $\chi$, even if $\chi \ll \kappa$. This result may seem odd, given that it implies a finite measurement rate as $\chi \rightarrow 0$. However, we consider the situation where as $\chi$ decreases the squeezing increases to compensate, such that the measurement rate remains finite.

To understand the significance of this result, we compare it against the measurement rate of a standard dispersive measurement made with the {\it same} total number of intracavity photons, but where the qubit coupling is increased to its optimal value $\chi \rightarrow \kappa/2$.  We find:
\begin{align}
	\frac{ \Gamma_{\rm ISTMS}[\chi \ll \kappa] }{\Gamma_{\rm SM, opt}[\chi = \kappa/2]} =
		\frac{\bar{n}_0}{\bar{n}_0 + \frac{\kappa}{4 \chi} }
	\label{eq:SNRWeakRatio}
\end{align}
If the ISTMS measurement is made with
a sufficiently large coherent drive (such that that $\bar{n}_0 \gg \kappa / \chi$), the measurement rate for a weakly coupled qubit can match the maximum measurement rate of a standard dispersive measurement (something that necessarily requires strong coupling).

Squeezing in the ISTMS scheme thus lets one in principle completely overcome the limitation of having a weakly coupled qubit without having to compensate with an extremely large number of measurement photons.  This is also accomplished without having to transport and inject an externally prepared squeezed state into a cavity with high efficiency.
We stress that this optimal scaling is not possible in a standard single-cavity setup, even if one could perfectly inject externally-generated
single-mode squeezing; this is because
the interaction with the qubit always mixes squeezed and anti-squeezed quadratures  \cite{Barzanjeh2014,Didier2015}, something that does not happen in our scheme (c.f.~Fig.~\ref{fig:SqzSpec}).

%%%%%%%%%%%%%%%%%%%%%%%%%%%%
%%%%%%%%%%%%%%%%%%%%%%%%%%%%

%%%%%%%%%%%%%%%%%%%%%%%%
\section{Short-time measurement dynamics}

While our discussion so far focused on the long-time properties of a measurement (as described by the measurement rate), the ISTMS setup
also provides strong advantages for finite measurement times.  We quantify the speed of readout by the measurement time required to reach $99.99\%$ fidelity, $\tau^*$. The fidelity is given by $F = 1- {\rm erfc}({\rm SNR}/2)/2$, where SNR denotes the signal-to-noise ratio.
We will be interested in situations where $\tau^*  \kappa$ is finite, and thus use in what follows the full expression for the time-dependent SNR (which reflects the non-white output noise spectrum) found in \ref{sec:inout}.

\begin{figure}[h!]
\subfigure{
\includegraphics[width=0.8\columnwidth]{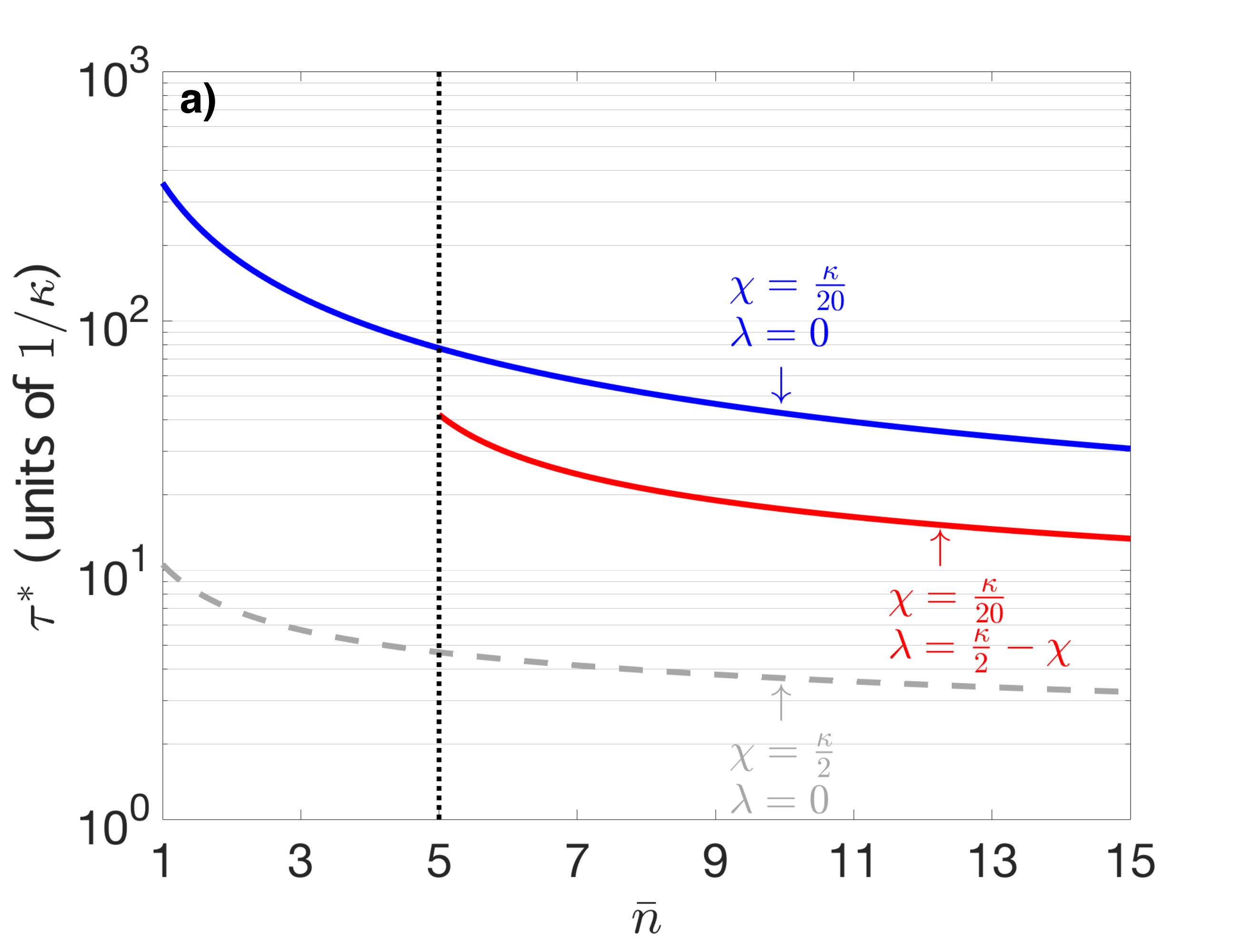}
\label{fig:MeasTimeA}}
\subfigure{
\includegraphics[width=0.8\columnwidth]{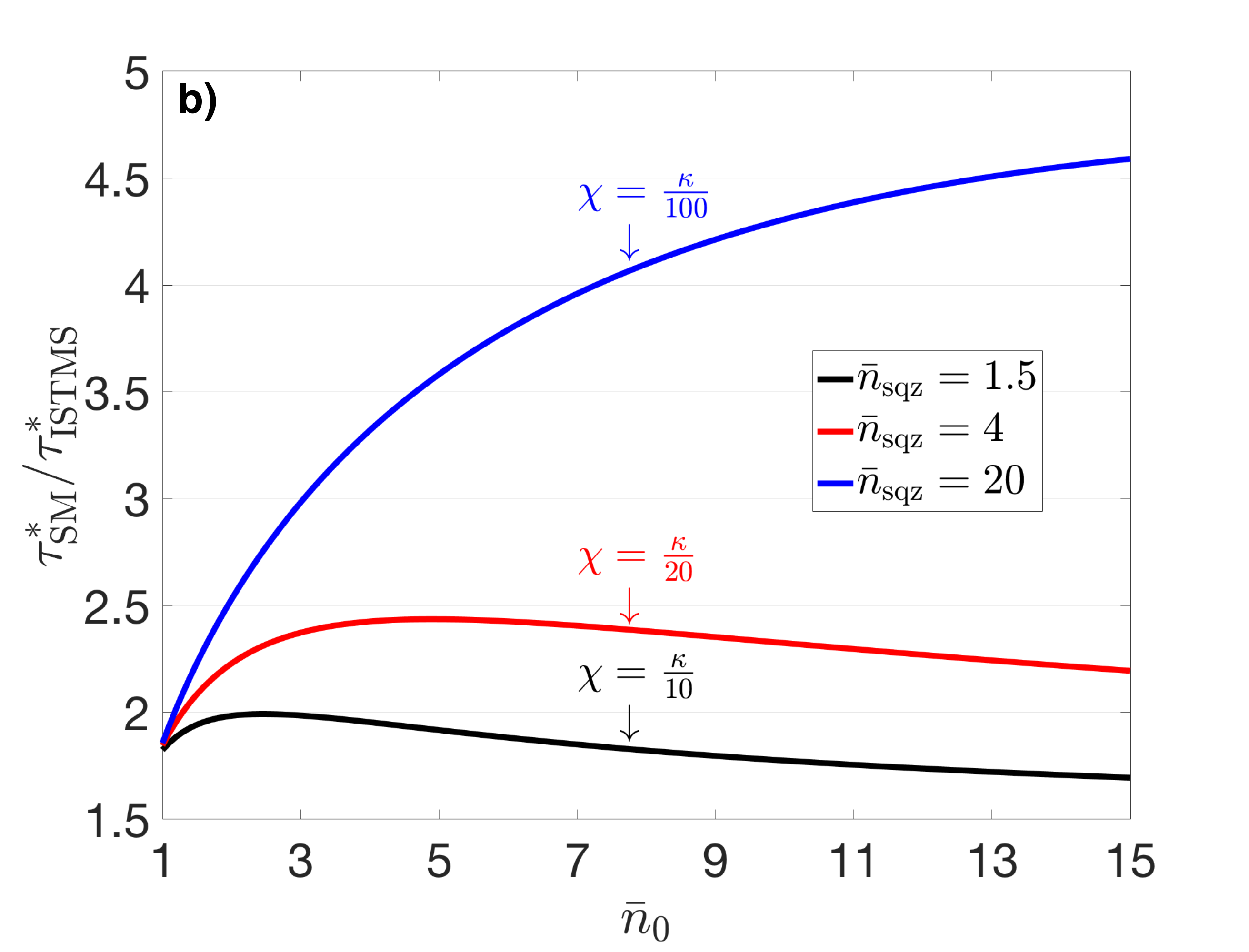}
\label{fig:MeasTimeB}}
\caption{{\bf a)} Measurement time $\tau^*$ needed for a fidelity $F=0.9999$ as a function of total intracavity photon number. Solid curves correspond to a weak qubit-cavity coupling $\chi = \kappa/20$, while the dashed curve describes an optimal strong coupling $\chi = \kappa/2$.  For the ISTMS setup (red curve) the number of squeezing photons $n_{\rm sqz} \approx 4$, indicated by the dotted vertical line. {\bf b)} Ratio of $\tau^*$ for standard readout with zero squeezing to that for the ISTMS setup with $\la = \kk/2-\chi$. $\tau^*$ can be reduced almost five-fold for the weakest coupling.}
\label{fig:MeasTime}
\end{figure}

We find that for weak couplings $\chi \ll \kappa$, the qubit-induced squeezing disruption is minimal, and the ISTMS scheme results in substantial reductions of $\tau^*$ compared to a standard dispersive measurement made with the same coupling and average intracavity photon number.  This is shown explicitly in Fig.~\ref{fig:MeasTime}.  For weak couplings $\chi = 10^{-2} \kappa$, the ISTMS scheme results in an almost five-fold reduction in $\tau^*$ in the limit of large $\bar{n}_0$.

Despite the large $\bar{n}_0$, the total intracavity photon number is below the critical photon number, which is given by $n_{\rm crit} = J/4\chi$ for the ISTMS setup. For a realistic $J = 10\kk$, this gives $n_{\rm crit} = 25,\ 50,\ 250$, for $\kk/\chi = 10,\ 20,\ 100$. As we are below $n_{\rm crit}$, we do not expect to see the effects of higher-order nonlinearities arising from the full Jaynes-Cummings interaction, and furthermore, there is a mounting body of evidence that even as one approaches $n_{\rm crit}$, the state of the system is not greatly affected \cite{Jeffrey:2014zr,Khezri:2016aa,Sank:2016aa}. In support of this intuition, a numerical comparison to the full Jaynes-Cummings model is discussed in \ref{sec:ApJC}.

%%%%%%%%%%%%%%%%%%
%%%%%%%%%%%%%%%%%%

%%%%%%%%%%%%%%%%%%
%%%%%%%%%%%%%%%%%%
\section{Effect of losses}

While the ISTMS setup does not suffer from the problems associated with transporting and injecting a squeezed state, its efficacy will be
degraded if there are significant internal cavity losses $\kappa_{\rm int}$.
Taking both cavities to have the same level of internal loss, and
letting $\varepsilon = \kappa_{\rm int} / ({\kappa_{\rm int} + \kappa})$,
we find that the weak-coupling measurement rate in Eq.~(\ref{eq:SNRWeak}) (which is independent of coupling) is not modified by internal loss as long as $(\chi/\kk)^2 > \varepsilon/2$; in this regime the reductions in measurement time depicted in Fig.~\ref{fig:MeasTime} are also not significantly impacted.  Further details on the effects of internal losses (and external measurement efficiencies) are presented in \ref{sec:ApLoss}.

\section{Suppression of Purcell decay and other extraneous qubit backaction}
\label{sec:PDSup}

An obvious concern with the ISTMS setup is that anti-squeezing generated by the parametric drive will greatly enhance non-QND backaction effects on the qubit.  In a standard dispersive setup, the non-QND backaction corresponds to Purcell decay, the cavity-enhanced relaxation of the qubit excited state \cite{Blais2004}.  In our setup, one might expect that the parametric driving enhances both the cavity-induced qubit decay and qubit heating.

\begin{figure}[h!]
\subfigure{
\includegraphics[width=0.45\columnwidth]{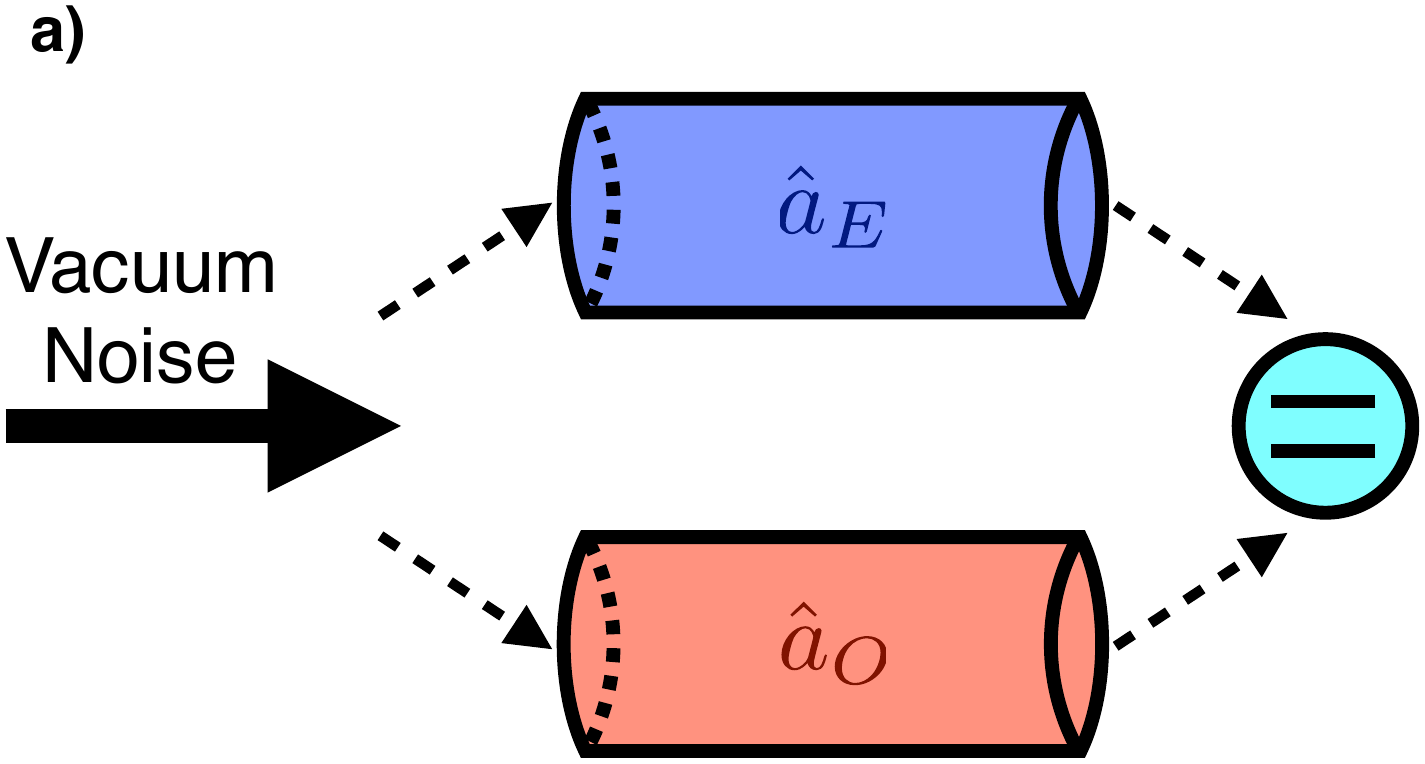}
\label{fig:Schem}}
\subfigure{
\includegraphics[width=0.45\columnwidth]{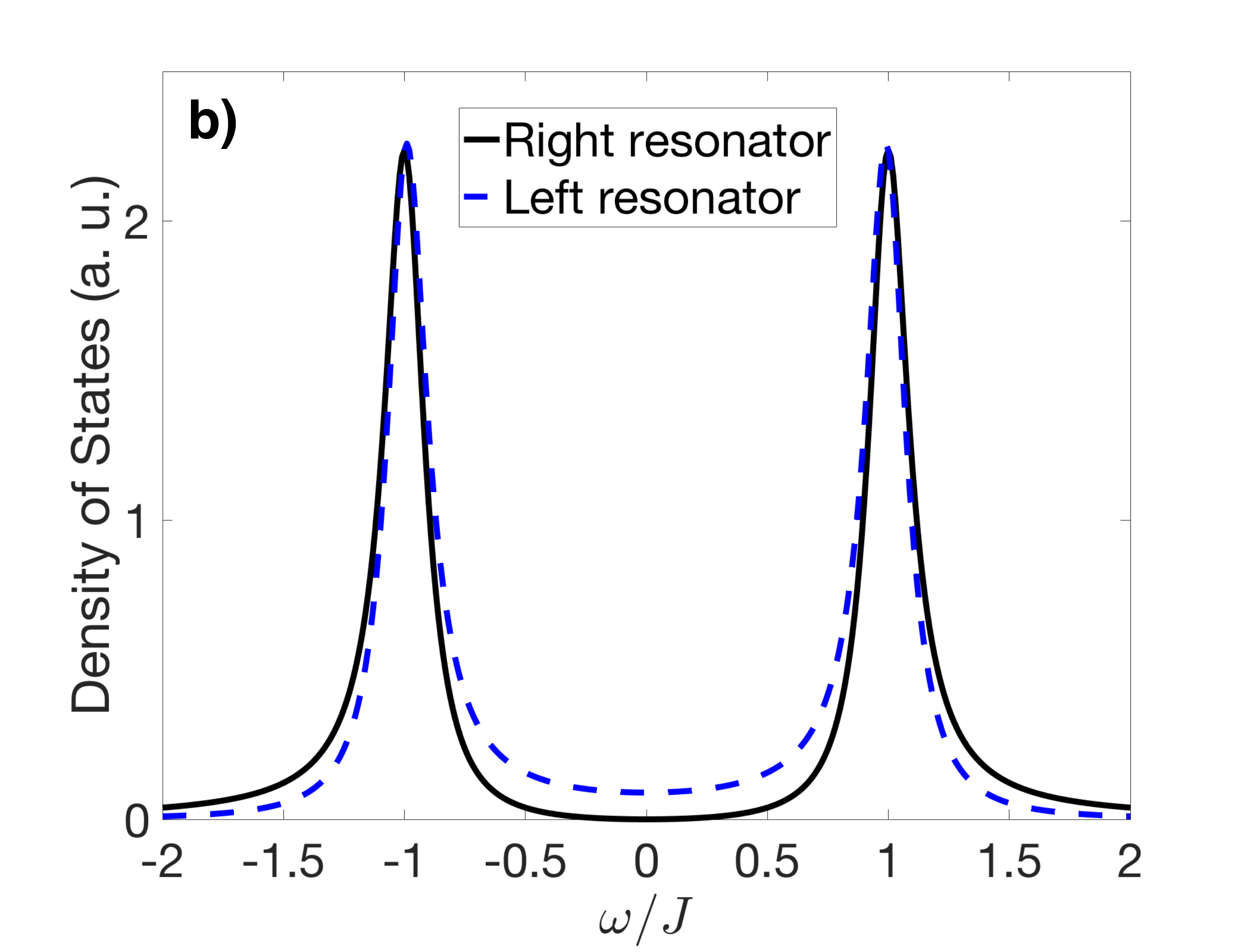}
\label{fig:DoS}}
\caption{{\bf a)} Schematic of the destructive interference leading to cancelation of the vacuum noise at the qubit frequency. {\bf b)} Density of states (see \ref{sec:ApBA}) of the right and left cavities for $\la = 0$ and $J = 5\kk$. Due to quantum noise interference the DoS of the right cavity is exactly zero at the qubit frequency ($\omega = 0$ in this frame).}
\end{figure}

Remarkably, these are not issues in the ISTMS setup.  This is the result of an interference effect involving the two cavities in the setup that cause it to act as its own, intrinsic Purcell filter, thus removing the necessity to construct an external filter \cite{Reed2010,Jeffrey:2014zr,Govia:2015aa,Bronn:2015aa}.  Heuristically, vacuum noise entering the right cavity can either drive the even or odd cavity normal mode before reaching the qubit (see Fig.~\ref{fig:Schem}).  The interference between these possibilities is perfectly destructive at the qubit frequency (which is at the
average to the two cavity normal mode frequencies).  Formally, this results in a suppression of the effective photonic density of states seen by the qubit (see Fig.~\ref{fig:DoS} and \ref{sec:ApBA}).

%%%%%%%%%%%%%%
%%%%%%%%%%%%%%
Alternatively, one can understand the effect by considering how the cavity normal modes transform under the dispersive transformation.
One finds to leading order in $1/J$
\begin{align}
	\hat{\tilde{a}}_{E/O}  \simeq \hat{a}_{E/O} + \frac{g}{\sqrt{2}J}\left(\mp\hat{\sigma}^- + \frac{\la}{J}\hat{\sigma}^+\right),
\end{align}
where $\hat{\tilde{a}}_{E/O}$ denotes the transformed version of the original normal mode lowering operators $\hat{a}_{E/O}$ .  As is standard, the weak hybridization of the qubit and cavity modes cause $\hat{\tilde{a}}_{E/O}$ to acquire a term involving the qubit lowering operator $\hat{\sigma}^-$.  More unusual is the fact that these operators also mix with the qubit raising operator $\hat{\sigma}^+$ due to the parametric drive $\lambda$.

The transmission line coupled to the right cavity couples to $\hat{a}_R$ via a system-bath coupling Hamiltonian of the form
$\hat{H}_{SB} = \sqrt{2\kappa} \hat{B}^\dagger \hat{a}_R + h.c. $, where $\hat{B}^\dagger$ is a bath operator that creates an excitation in the transmission
line (see, e.g., \cite{Clerk:2010aa}).  In the dispersive frame, $\hat{a}_R$ becomes:
\begin{align}
	\hat{\tilde{a}}_R  = \frac{\hat{\tilde{a}}_E + \hat{\tilde{a}}_O}{\sqrt{2}} \simeq \frac{\hat{a}_E + \hat{a}_O}{\sqrt{2}}+ \frac{g\la}{J^2}\hat{\sigma}^+.
\end{align}
Note that the $\hat{\sigma}^{-}$ perfectly cancels in this expression.  As a result, there is no amplitude for creating a bath excitation (via $\hat{B}^\dagger$) and relaxing the qubit, meaning that conventional Purcell decay is suppressed.

In contrast, there is a weak amplitude for both creating a bath excitation {\it and} exciting the qubit.
The corresponding heating rate of the qubit is readily found to be
$\yy_{H} = \kappa\left(g\la/J^2\right)^2$.
This heating is an example of ``quantum heating'', as discussed in Refs.~\cite{Dykman:2011aa,Marthaler:2006aa}.
While this heating is clearly unwanted, we stress that it is much weaker that standard Purcell decay in the regime of interest:  standard Purcell decay would scale as $(g / J)^2 \kappa$, where as the above heating effect is suppressed by an additional extremely small parameter $(\lambda / J)^2 < (\kappa / 2J)^2$, with $(\kappa / 2J)^2 \sim 1/100$ achievable for realistic values of $J$ \cite{Eichler:2014aa}. Qubit dissipation effects due to internal loss in the left cavity are treated in \ref{sec:ApBA}.

%%%%%%%%%%%%%%%%
\section{Conclusion}

We have presented an analysis of a realistic two-cavity setup that allows squeezing to dramatically enhance the readout of weakly-coupled qubits:  our setup essentially allows one to recover the optimal measurement rate of a strongly-coupled system while using a weakly-coupled system.  This is accomplished in a manner that does not require one to transport and inject an externally-produce squeezed state with high fidelity.  The system is also remarkably robust against non-QND backaction effects, due to an interference effect that causes the system to act as its own, intrinsic Purcell filter.  There are many possible routes for realizing the parametric drive used in our setup.  One could flux-pump a SQUID embedded in the left cavity  \cite{Yamamoto:2008aa}, or use a Josephson-junction mediated coupling to an external pump mode, as has been done to great effect in Josephson parametric converter circuits \cite{Bergeal:2010aa,Bergeal2010,Sliwa:2015aa}.  Finally, while the focus of our work has been on circuit/cavity QED, the ISTMS setup and the results obtained here are completely general, and can be applied to other physical systems, such as optomechanics \cite{Peano:2015aa}.

\section{Acknowledgements}

This work was supported by the Army Research Office under Grant No. W911NF-14-1-0078.

%%%%%%%%%%%%%%%%%%%%%%%%%%%%%%%%%%%%%%%%%%%%%%%%%%%%%%%%%%
%%%%%%%%%%%%%%%%%%%%%%%%%%%%%%%%%%%%%%%%%%%%%%%%%%%%%%%%%%

\section*{Appendices}

\appendix

\section{Two-Mode Squeezing}
\label{sec:AppTM}

We begin with the Hamiltonian describing two tunnel-coupled cavities, one of which has an embedded SQUID that is flux-pumped at frequency $\ww_p$  \cite{Yamamoto:2008aa}
\begin{align}
 \nonumber\hat{H} &= \ww_c \hat{a}_L^\dagger{a}_L + \ww_c \hat{a}_R^\dagger{a}_R + J\left(\hat{a}_L^\dagger\hat{a}_R+\hat{a}_R^\dagger\hat{a}_L\right) \\ &+ i\lambda\left(e^{i\ww_p t} - e^{-i\ww_p t}\right)\left(\hat{a}_L - \hat{a}_L^\dagger\right)^2,
\end{align}
where $J$ is the tunnel coupling strength and $\la$ is the strength of the flux pump. We define the eigenmodes of the coupled cavities as
\begin{align}
\hat{a}_E = \frac{\hat{a}_L+\hat{a}_R}{\sqrt{2}},\ \ \hat{a}_O = \frac{\hat{a}_R-\hat{a}_L}{\sqrt{2}}.
\end{align}
In this basis, the Hamiltonian is given by
\begin{align}
  \nonumber\hat{H} &= \ww_E \hat{a}_E^\dagger{a}_E + \ww_O \hat{a}_O^\dagger{a}_O \\ &+ \frac{i\lambda}{2}\left(e^{i\ww_p t} - e^{-i\ww_p t}\right)\left(\hat{a}_E  - \hat{a}_O  - \hat{a}_E^\dagger + \hat{a}_O^\dagger\right)^2,
\end{align}
where $\ww_E = \ww_c + J$ and $\ww_O = \ww_c -J$.
Defining $\hat{A} = \left(\hat{a}_E  - \hat{a}_O  - \hat{a}_E^\dagger + \hat{a}_O^\dagger\right)^2$ and expanding we have
\begin{align}
\nonumber\hat{A} = \hat{a}_E\hat{a}_E +  \hat{a}_O\hat{a}_O + \hat{a}^\dagger_E\hat{a}^\dagger_E + \hat{a}^\dagger_O\hat{a}^\dagger_O -2 \hat{a}_E\hat{a}_O \\ -2 \hat{a}^\dagger_E\hat{a}^\dagger_O - 2\hat{a}^\dagger_E\hat{a}_E -2\hat{a}^\dagger_O\hat{a}_O +2\hat{a}^\dagger_O\hat{a}_E +2\hat{a}^\dagger_E\hat{a}_O,
\end{align}
and in the interaction frame this becomes
\begin{align}
\nonumber&\hat{A}_I = \ e^{-2i\ww_Et}\hat{a}_E\hat{a}_E +  e^{-2i\ww_Ot}\hat{a}_O\hat{a}_O + e^{2i\ww_Et}\hat{a}^\dagger_E\hat{a}^\dagger_E \\  \nonumber&+ e^{2i\ww_Ot}\hat{a}^\dagger_O\hat{a}^\dagger_O - 2\hat{a}^\dagger_E\hat{a}_E -2\hat{a}^\dagger_O\hat{a}_O \\
\nonumber&-2e^{-i\left(\ww_E+\ww_O\right)t} \hat{a}_E\hat{a}_O -2e^{i\left(\ww_E+\ww_O\right)t} \hat{a}^\dagger_E\hat{a}^\dagger_O \\
& +2e^{-i\left(\ww_E-\ww_O\right)t} \hat{a}^\dagger_O\hat{a}_E +2e^{i\left(\ww_E-\ww_O\right)t} \hat{a}^\dagger_E\hat{a}_O.
\end{align}
Setting $\ww_p = 2\ww_c$, we drop all terms that oscillate rapidly at frequencies on the order of $\ww_c \gg J$ in the product $\left(e^{i\ww_p t} - e^{-i\ww_p t}\right)\hat{A}_I$ to arrive at the Hamiltonian in the interaction frame
\begin{align}
\nonumber\hat{H}_I &= -i\la\left(\hat{a}_E\hat{a}_O-\hat{a}^\dagger_E\hat{a}^\dagger_O\right) \\ &+\frac{i\la}{2}\left(e^{-i2Jt}\hat{a}_E\hat{a}_E+e^{i2Jt}\hat{a}_O\hat{a}_O- h.c.\right).
\end{align}
If $\la/(4J) \ll 1$, then we can also drop the single mode squeezing terms to arrive at the final Hamiltonian in the interaction frame
\begin{align}
\hat{H}_I = -i\la\left(\hat{a}_E\hat{a}_O-\hat{a}^\dagger_E\hat{a}^\dagger_O\right),
\end{align}
and in the lab frame
\begin{align}
\nonumber\hat{H} &=  \ww_E \hat{a}_E^\dagger{a}_E + \ww_O \hat{a}_O^\dagger{a}_O\\&-i\la\left(e^{i2\ww_c t}\hat{a}_E\hat{a}_O-e^{-i2\ww_c t}\hat{a}^\dagger_E\hat{a}^\dagger_O\right). \label{eqn:LabH}
\end{align}

\section{Dispersive Approximation}
\label{sec:ApDis}

We now include coupling between the right cavity and the qubit to the two-mode squeezing Hamiltonian of Eq.~(\ref{eqn:LabH}). In the eigenmode basis and in a frame rotating at $\ww_c = \ww_q = \ww_p/2$ for both eigenmodes and the qubit, the Hamiltonian is given by
\begin{align}
\nonumber&\hat{H}' = J \hat{a}_E^\dagger{a}_E -J \hat{a}_O^\dagger{a}_O-i\la\hat{a}_E\hat{a}_O+i\la\hat{a}^\dagger_E\hat{a}^\dagger_O \\ &+\frac{g}{\sqrt{2}}\left(\hat{a}_E^\dagger\hat{\sigma}^- + \hat{\sigma}^+\hat{a}_E\right)+\frac{g}{\sqrt{2}}\left(\hat{a}_O^\dagger\hat{\sigma}^- + \hat{\sigma}^+\hat{a}_O\right). \label{eqn:HamJC}
\end{align}
We make a Schrieffer-Wolff transformation via the generator $S = S_E + S_O$,
where
\begin{align}
 &S_E = u\hat{a}_E\hat{\sigma}^+ + v\hat{a}_E\hat{\sigma}^- - u^*\hat{a}_E^\dagger\hat{\sigma}^- - v^*\hat{a}_E^\dagger\hat{\sigma}^+, \\
 &S_O = n\hat{a}_O\hat{\sigma}^+ + m\hat{a}_O\hat{\sigma}^- - n^*\hat{a}_O^\dagger\hat{\sigma}^- - m^*\hat{a}_O^\dagger\hat{\sigma}^+,
\end{align}
with
\begin{align}
&u = -n = -\frac{g}{\sqrt{2}J}\left(1+\frac{\la^2}{J^2}\right)^{-1},  \\
&v = m = \frac{ig\la}{\sqrt{2}J^2}\left(1+\frac{\la^2}{J^2}\right)^{-1},
\end{align}
chosen to cancel terms of order $g$ in the transformed Hamiltonian. Keeping only terms up to second order in $g/J$, we obtain the dispersive Hamiltonian
\begin{align}
\nonumber\hat{H}_D &= J \hat{a}_E^\dagger{a}_E - J \hat{a}_O^\dagger{a}_O +\chi\left(\hat{a}_E^\dagger{a}_E - \hat{a}_O^\dagger{a}_O\right)\hat{\sigma}_z \\ \nonumber&- i\la\left(1-\frac{\chi}{J}\hat{\sigma}_z\right)\left(\hat{a}_E\hat{a}_O - \hat{a}^\dagger_E\hat{a}^\dagger_O\right) \\
\nonumber&+ \frac{i\la\chi}{2J}\left(1-\frac{\chi}{J}\hat{\sigma}_z\right)\left(\hat{a}_E^2-(\hat{a}_E^\dagger)^2\right)\hat{\sigma}_z \\ &+  \frac{i\la\chi}{2J}\left(1-\frac{\chi}{J}\hat{\sigma}_z\right)\left(\hat{a}_O^2-(\hat{a}_O^\dagger)^2\right)\hat{\sigma}_z. \label{eqn:HamAD}
\end{align}
where
\begin{align}
\chi =\frac{g^2}{2J}\left(1+\frac{\la^2}{J^2}\right)^{-1},
\end{align}
which we approximate as $\chi =g^2/2J$ in the main text as $\la/J$ is a small parameter. As a result of the fact that the qubit frequency is placed exactly midway between the two cavity eigenmode frequencies, there is no second order qubit-mediated eigenmode coupling due to destructive interference.

The terms in the last two lines of Eq.~(\ref{eqn:HamAD}) are single mode squeezing terms for each eigenmode. However, these terms are both off resonance and are damped by the small parameter $\chi/2J$. Therefore, provided $\la\chi/(4J^2) \ll 1$ they can be safely ignored, and one obtains the dispersive Hamiltonian used in the main text.

\section{Measurement Dynamics and Signal-to-Noise Ratio}
\label{sec:inout}

Starting from the Hamiltonian in the dispersive frame (Eq.~(2) of the main text) we derive the equations of motion for the even and odd modes
\begin{align}
&\dot{\hat{a}}_E =  -i\chi\hat{\sigma}_z\hat{a}_E +\lambda\hat{a}_O^\dagger - \frac{\kk}{2}\hat{a}_E - \sqrt{\kk}\hat{b}_{E, {\rm in}}(t), \label{eqn:aE} \\
&\dot{\hat{a}}_O = i\chi\hat{\sigma}_z\hat{a}_O +\lambda\hat{a}_E^\dagger - \frac{\kk}{2}\hat{a}_O - \sqrt{\kk}\hat{b}_{O, {\rm in}}(t), \label{eqn:aO}
\end{align}
where $\kk$ is the decay rate of the eigenmodes into the transmission line, and $\hat{b}_{E/O, {\rm in}}(t)$ are the input fields for each eigenmode. The above equations are obtained in the limit $J\gg\kk$ where the environments of the even and odd modes can be treated as independent, as the even and odd modes effectively couple to orthogonal collections of modes in the right transmission line due to their large frequency separation, $2J$, and small linewidth, $\kk$. For the purpose of calculating the SNR we need only consider the qubit in a definite state, and as such we will write $\hat{\sigma}_z$ as a real number $\sigma = \pm1$ from here on.

In terms of the quadratures of the plus and minus modes, the system evolution equations are
\begin{align}
&\dot{\hat{X}}_+ = \chi\sigma\hat{Y}_- + (\la-\frac{\kk}{2})\hat{X}_+ - \sqrt{\kk}\hat{X}_{+, {\rm in}}(t), \\
&\dot{\hat{X}}_- = \chi\sigma\hat{Y}_+ - (\la+\frac{\kk}{2})\hat{X}_- - \sqrt{\kk}\hat{X}_{-, {\rm in}}(t), \label{eqn:sqX}\\
&\dot{\hat{Y}}_+ = -\chi\sigma\hat{X}_- - (\la+\frac{\kk}{2})\hat{Y}_+ - \sqrt{\kk}\hat{Y}_{+, {\rm in}}(t), \label{eqn:sqY} \\
&\dot{\hat{Y}}_- = -\chi\sigma\hat{X}_+ + (\la-\frac{\kk}{2})\hat{Y}_- - \sqrt{\kk}\hat{Y}_{-, {\rm in}}(t),
\end{align}
with the input quadratures defined appropriately. We will focus on the squeezed quadratures $\hat{X}_-$ and $\hat{Y}_+$, whose evolution is decoupled from that of the amplified quadratures. In frequency space, the solutions for the squeezed quadratures are given by
\begin{widetext}
\begin{align}
\left(\begin{array}{c}
\hat{X}_-(\ww) \\
\hat{Y}_+(\ww)
\end{array}\right) =
\frac{-\sqrt{\kk}}{(\la + \frac{\kk}{2} + i\ww)^2+\chi^2}\left(\begin{array}{cc}
i\ww + \la + \frac{\kk}{2} & -\chi\sigma \\
\chi\sigma & i\ww + \la + \frac{\kk}{2}
\end{array}\right)
\left(\begin{array}{c}
\hat{X}_{-,{\rm in}}(\ww) \\
\hat{Y}_{+,{\rm in}}(\ww)
\end{array}\right)
\label{eqn:Solutions}
\end{align}
\end{widetext}
and using the usual input-output relation $\hat{X}_{-, {\rm out}}(\ww) = \hat{X}_{-, {\rm in}}(\ww) + \sqrt{\kk}\hat{X}_-(\ww)$, we can calculate the output field of the $\hat{X}_-$ quadrature in frequency space
\begin{align}
\nonumber\hat{X}_{-, {\rm out}}(\ww) &= \left(1-\frac{\kk(\la + \frac{\kk}{2} + i\ww)}{(\la + \frac{\kk}{2} + i\ww)^2+\chi^2}\right)\hat{X}_{-, {\rm in}}(\ww) \\&+ \frac{\kk\chi\sigma}{(\la + \frac{\kk}{2} + i\ww)^2+\chi^2}\hat{Y}_{+, {\rm in}}(\ww). \label{eqn:Xout}
\end{align}

\subsection{Measurement Protocol}

We consider a homodyne measurement, for which the homodyne current will be proportional to the average value of the output quadrature $\hat{X}_{-,{\rm out}}(t)$. It is important to note that while the measurement signal is contained in only one quadrature, $\hat{X}_{-, {\rm out}}(t)$, this quadrature exists in a frame rotating at different frequencies for the two cavities, and cannot be directly measured. The physical measurement requires measuring the homodyne currents for both the even and odd modes, i.e. $\left<\hat{X}_{E, {\rm out}}(t)\right>$ and $\left<\hat{X}_{O,{\rm out}}(t)\right>$, which are separated in frequency by $2J$. Calculation of $\left<\hat{X}_{-,{\rm out}}(t)\right>$ can then be done via classical post processing of these signals. As input we consider $\left<\hat{Y}_{+,{\rm in}}(t)\right> \neq 0$ and $\left<\hat{X}_{-,{\rm in}}(t)\right>= 0$, such that in light of Eq.~(\ref{eqn:Xout}), $\left<\hat{X}_{-,{\rm out}}(t)\right>$ will be qubit-state dependent. As was the case for the measurement quadrature, the input quadrature $\hat{Y}_{+,{\rm in}}(t)$ cannot be directly accessed, so the physical input will consist of coherently driving both input modes, $\hat{Y}_{E,{\rm in}}(t)$ and $\hat{Y}_{O,{\rm in}}(t)$, such that the correct coherent drive is applied to the combined quadrature $\hat{Y}_{+,{\rm in}}(t)$.

\subsection{Measurement Signal}

We set
\begin{align}
\left<\hat{Y}_{+,{\rm in}}(t)\right> = \sqrt{2}\abs{\beta}\Theta(t-t_0),
\end{align}
which describes a coherent pulse incident on the system beginning at time $t_0$, which we assume is long after the system has reached its undriven steady-state. Without loss of generality we can set $t_0 = 0$. The factor of $\sqrt{2}$ comes from the fact that we define the input field of the plus mode as  $\left<\hat{b}_{+,in}(t)\right> = i\abs{\beta} \Theta(t-t_0)$, so that the input is entirely in the $\hat{Y}_+$ quadrature. For this choice of input field, the expectation value of the output field is given by
\begin{align}
\nonumber&\left<\hat{X}^{\sigma}_{-, {\rm out}}(t)\right> = \frac{-\sqrt{2}\kk\chi\abs{\beta}\sigma}{\left(\la+\frac{\kk}{2}\right)^2+\chi^2}\Bigg[1 \\&- \left(\cos\left(\chi t\right)+\frac{\la+ \kk/2}{\chi}\sin\left(\chi t\right)\right)e^{-\left(\la+\frac{\kk}{2}\right)t}\Bigg]. \label{eqn:Mop}
\end{align}

We define the measurement operator as $\hat{M}(t) = \hat{X}_{-, {\rm out}}^{+1}(t) - \hat{X}^{-1}_{-, {\rm out}}(t)$, and it is clear from Eq.~(\ref{eqn:Mop}) that the expectation value of this operator is $\left<\hat{M}(t)\right> = 2\left<\hat{X}_{-, {\rm out}}^{+1}(t)\right>$. Using this, we calculate the integrated measurement signal
\begin{widetext}
\begin{align}
\nonumber&\mathcal{M}_S(\tau) = \sqrt{\kk}\int_0^\tau dt \left<\hat{M}(t)\right> = \\ &\frac{-2\sqrt{2\bar{n}_0}\chi}{\sqrt{\left(\la+\frac{\kk}{2}\right)^2+\chi^2}}\Bigg[\kk\tau-\frac{\kk e^{-\left(\la+\frac{\kk}{2}\right)\tau}}{\left(\la+\frac{\kk}{2}\right)^2+\chi^2}\Bigg(\frac{\chi^2-(\la+\frac{\kk}{2})^2}{\chi}\sin\left(\chi\tau\right)-2\left(\la+\frac{\kk}{2}\right)\left[\cos\left(\chi\tau\right) - e^{\left(\la+\frac{\kk}{2}\right)\tau}\right]\Bigg)\Bigg],
\end{align}
\end{widetext}
where we have reparameterized in terms of the coherent intracavity photon number $\bar{n}_0$, rather than the input photon flux $\abs{\beta}$. In the asymptotic long-time limit, $\kk\tau \gg 1$, the measurement signal can be approximated by
\begin{align}
\mathcal{M}_S(\tau) \rightarrow \frac{-2\sqrt{2\bar{n}_0}\chi}{\sqrt{\left(\la+\frac{\kk}{2}\right)^2+\chi^2}} \kk\tau.
\end{align}

\subsection{Noise}

Using Eq.~(\ref{eqn:Xout}) we calculate the output noise to be
\begin{align}
\nonumber &M^{\sigma}_N(t,t') = \left<\hat{X}^{\sigma}_{-,{\rm out}}(t)\hat{X}^{\sigma}_{-,{\rm out}}(t')\right> = \frac{1}{2}\bigg(\delta(t-t') \\ \nonumber&-\frac{\kk\la}{2(\la+\frac{\kk}{2})}\Big[\Theta(t-t')e^{-\left(\la+\frac{\kk}{2}\right)(t-t')}\left(e^{i\chi(t-t')}+e^{-i\chi(t-t')}\right)\\
&+\Theta(t'-t)e^{-\left(\la+\frac{\kk}{2}\right)(t'-t)}\left(e^{i\chi(t'-t)}+e^{-i\chi(t'-t)}\right)\Big]\bigg),
\end{align}
where $\Theta(t-t')$ is the Heaviside step function. As expected, the noise is independent of the state of the qubit, and we can suppress the upper index $\sigma$. With this expression we can calculate the integrated noise
\begin{widetext}
\begin{align}
\nonumber&\mathcal{M}_N(\tau) = \kappa\int_0^\tau dt \int_0^\tau dt' M_N(t,t') =\frac{1}{2}\left(1-\frac{2\kk\la}{\chi^2+\left(\la+\frac{\kk}{2}\right)^2}\right)\kk\tau \\&- \frac{\kk^2\la}{\left(\la+\frac{\kk}{2}\right)\left[\chi^2+\left(\la+\frac{\kk}{2}\right)^2\right]^2}\Bigg[\left(\chi^2-\left(\la+\frac{\kk}{2}\right)^2\right)\left(1-\cos\left(\chi\tau\right)e^{-\left(\la+\frac{\kk}{2}\right)\tau}\right)
-2\chi\left(\la+\frac{\kk}{2}\right)\sin\left(\chi\tau\right)e^{-\left(\la+\frac{\kk}{2}\right)\tau}\Bigg],
\end{align}
\end{widetext}
and in the asymptotic long-time limit this reduces to
\begin{align}
\mathcal{M}_N(\tau) \rightarrow \frac{1}{2}\frac{\chi^2+\left(\la-\frac{\kk}{2}\right)^2}{\chi^2+\left(\la+\frac{\kk}{2}\right)^2} \kk\tau.
\end{align}

\subsection{Squeezing Spectrum}

The squeezing spectrum is defined by
\begin{align}
S_{{\rm out}}(\omega) = \int_0^\infty d\ww' \left<\hat{X}_{-,{\rm out}}(\ww)\hat{X}_{-,{\rm out}}(\ww')\right>
\end{align}
where $\hat{X}_{-,{\rm out}}(\ww)$ is the Fourier transform of $\hat{X}_{-,{\rm out}}(t)$. For the ISTMS setup, the squeezing spectrum is given by
\begin{align}
 \nonumber S_{\rm out}(\ww) &= \frac{1}{2}\Bigg(1 - \frac{\kk\la}{\la+\frac{\kk}{2}}\Bigg[\frac{\la+\frac{\kk}{2}}{\left(\la+\frac{\kk}{2}\right)^2+\left(\ww+\chi\right)^2}\\&+\frac{\la+\frac{\kk}{2}}{\left(\la+\frac{\kk}{2}\right)^2+\left(\ww-\chi\right)^2}\Bigg]\Bigg),
\end{align}
which shows the double-Lorentzian shape seen in the main text. To measure the squeezing spectrum in dB, we compare to the squeezing spectrum for $\la = 0$, $S_{\rm out}(\ww)= 1/2$, using the formula
\begin{align}
  S_{\rm out}(\ww)\ (\rm dB) = 10\log_{10}\left[2S_{\rm out}(\ww)\right].
\end{align}

\subsection{Signal-to-Noise Ratio}

With both the integrated signal and the integrated noise, we can write down the signal to noise ratio
\begin{widetext}
\begin{align}
\nonumber{\rm SNR}^2 &= \frac{\left(\mathcal{M}_S(\tau)\right)^2}{\mathcal{M}^{+1}_N(\tau)+\mathcal{M}^{-1}_N(\tau)} = \frac{\left(\mathcal{M}_S(\tau)\right)^2}{2\mathcal{M}_N(\tau)}\\
&= \scriptstyle{\frac{\frac{8\bar{n}_0\chi^2}{\left(\la+\frac{\kk}{2}\right)^2+\chi^2}\left[\kk\tau
-\frac{\kk e^{-\left(\la+\frac{\kk}{2}\right)\tau}}{\left(\la+\frac{\kk}{2}\right)^2+\chi^2}\left(\frac{\chi^2-(\la+\frac{\kk}{2})^2}{\chi}\sin\left(\chi\tau\right)-2\left(\la+\frac{\kk}{2}\right)\cos\left(\chi\tau\right)\right)
-\frac{2\kk\left(\la+\frac{\kk}{2}\right)}{\left(\la+\frac{\kk}{2}\right)^2+\chi^2}\right]^2}{\frac{\chi^2+\left(\la-\frac{\kk}{2}\right)^2}{\chi^2+\left(\la+\frac{\kk}{2}\right)^2}\kk\tau - \frac{2\kk^2\la}{\left(\la+\frac{\kk}{2}\right)\left[\chi^2+\left(\la+\frac{\kk}{2}\right)^2\right]^2}\left[\left(\chi^2-\left(\la+\frac{\kk}{2}\right)^2\right)\left(1-\cos\left(\chi\tau\right)e^{-\left(\la+\frac{\kk}{2}\right)\tau}\right)-2\chi\left(\la+\frac{\kk}{2}\right)\sin\left(\chi\tau\right)e^{-\left(\la+\frac{\kk}{2}\right)\tau}\right]}}.
\end{align}
\end{widetext}
In the long-time limit, this becomes
\begin{align}
{\rm SNR}^2  \rightarrow \frac{8\bar{n}_0\chi^2}{\chi^2+\left(\la-\frac{\kk}{2}\right)^2}\kk\tau.
\end{align}

\section{Numerical Comparison of the Jaynes-Cummings and Dispersive Hamiltonians}
\label{sec:ApJC}

In order to quantify the effects of higher-order nonlinearities arising from interaction between the qubit and the cavities, we compare the steady-state of the dispersive Hamiltonian we have used in our calculations
\begin{align}
\nonumber\hat{H}_D &=  J \hat{a}_E^\dagger{a}_E - J \hat{a}_O^\dagger{a}_O +\chi\left(\hat{a}_E^\dagger{a}_E - \hat{a}_O^\dagger{a}_O\right)\hat{\sigma}_z \\&- i\la\left(\hat{a}_E\hat{a}_O - \hat{a}^\dagger_E\hat{a}^\dagger_O\right),
\end{align}
to that of the full Jaynes-Cummings model
\begin{align}
\nonumber\hat{H}' = J\left(\hat{a}_E^\dagger{a}_E - \hat{a}_O^\dagger{a}_O\right)-i\la\left(\hat{a}_E\hat{a}_O-\hat{a}^\dagger_E\hat{a}^\dagger_O\right) \\
+\frac{g}{\sqrt{2}}\left(\hat{a}_E^\dagger\hat{\sigma}^- + \hat{\sigma}^+\hat{a}_E + \hat{a}_O^\dagger\hat{\sigma}^- + \hat{\sigma}^+\hat{a}_O\right).
\end{align}
To do so, we numerically calculate the steady-states of the master equations
\begin{align}
	&\dot{\rho} = -i\left[\hat{H}_D,\rho\right] + \kk\mathcal{D}\left[\hat{a}_E\right]\rho + \kk\mathcal{D}\left[\hat{a}_O\right]\rho, \\
	&\dot{\rho} = -i\left[\hat{H}',\rho\right] + \kk\mathcal{D}\left[\hat{a}_E\right]\rho + \kk\mathcal{D}\left[\hat{a}_O\right]\rho,
\end{align}
where $\mathcal{D}\left[x\right]\rho = x\rho x^\dagger - \frac{1}{2}\left\{x^\dagger x, \rho\right\}$ is the usual dissipator. We have chosen to calculate the numerical steady-states in the even/odd basis, as in this basis the Jaynes-Cummings Hamiltonian can be made time-independent when expressed in a frame rotating at $\omega_c$. This makes numerical calculations less computationally intensive than if the Hamiltonian were time-dependent, as the Jaynes-Cummings Hamiltonian would be if expressed in the plus/minus basis. We compare the two steady-states by calculating their fidelity, defined as
\begin{equation}
	{\rm Fidelity} = {\rm Tr}\left(\sqrt{\sqrt{\rho_{D}}\rho_{JC}\sqrt{\rho_{D}}}\right).
\end{equation}

It is important to note that the Jaynes-Cummings and dispersive Hamiltonians are connected by a unitary transformation that ``mixes'' qubit and cavity observables, and as such the steady-states in the two frames are strictly speaking not directly comparable. However, based on previous work \cite{Govia:2016aa,Khezri:2016aa}, the scale of the lowest order effect of this mixing is set by $\chi/J$, and for our parameter choice ($\chi/J = 1/200$) this will have only a small effect on a direct comparison of the steady-states. Additionally, as we are primarily interested in the effect the Jaynes-Cummings dynamics has on the cavity state, we treat the even and odd mode decay as independent, and do not model the correlated decay discussed in Section \ref{sec:PDSup}. As such, in the Jaynes-Cummings simulation, there will be both Purcell decay and the quantum heating described in Section \ref{sec:PDSup}, which will lead to weak thermalization of the qubit steady-state. We have not included this effect in the dispersive simulation, and assumed that the qubit remains in its ground state. Given these effects, there will automatically be a small difference between the two steady-states we calculate.

\begin{figure}[h!]
\subfigure{
\includegraphics[width=0.8\columnwidth]{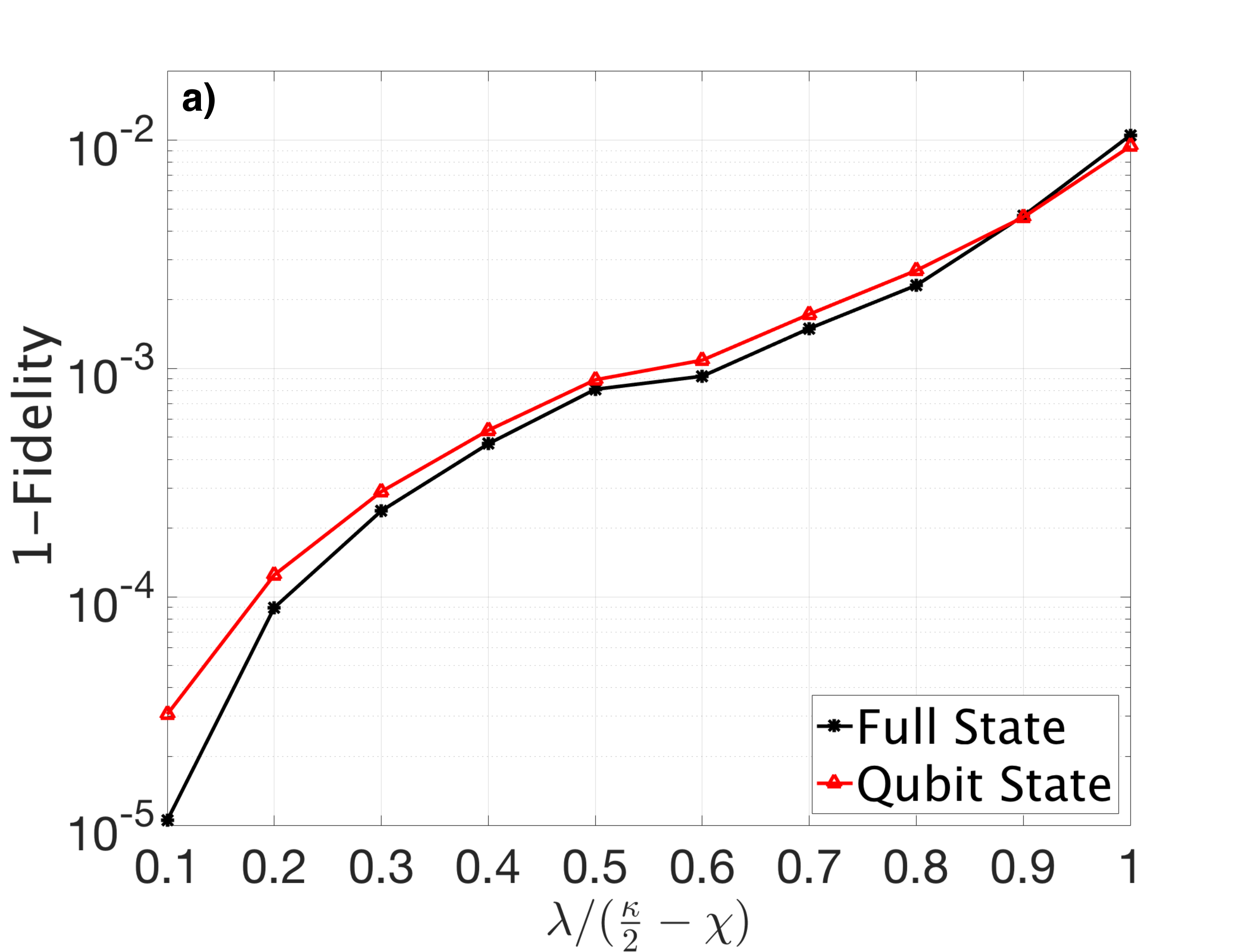}
\label{fig:JCcomp}}
\subfigure{
\includegraphics[width=0.8\columnwidth]{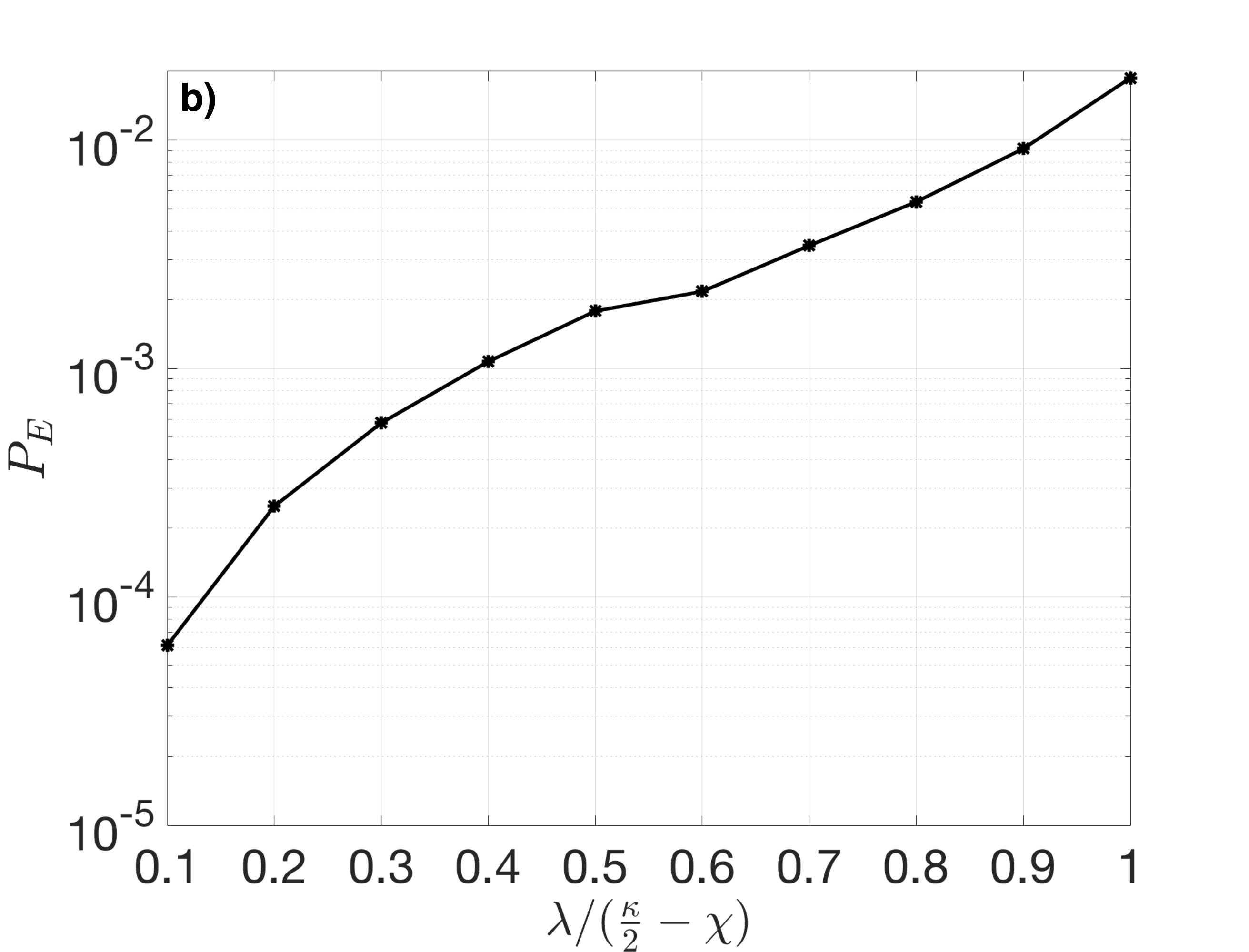}
\label{fig:EPop}}
\caption{{\bf a)} Error of the full steady-state (black stars) and qubit steady-state (red triangles) of the dispersive Hamiltonian compared to that for the full Jaynes-Cummings Hamiltonian, plotted as a function of the parametric drive strength. The system parameters are $\chi = \kappa/20$ and $\kappa = J/10$, which are well within experimental realization. {\bf b)} Qubit excited state population at steady-state for the Jaynes-Cummings Hamiltonian, as a function of the parametric drive strength.}
\end{figure}

Fig.~\ref{fig:JCcomp} shows that the fidelity error between the two steady-states grows as the parametric drive strength is increased. We measure the parametric drive strength in units of $\kk/2-\chi$, such that a value of one corresponds to an SNR that is independent of $\chi$, as in Eq.~(\ref{eq:SNRWeak}). This increase in error is to be expected, as an increased parametric drive strength leads to a larger intracavity photon number at steady-state, though we are still far below the critical photon number. Even for a large parametric drive strength the error is no more than $1\%$, which indicates that keeping only the lowest order in the dispersive approximation is justified for the ISTMS setup.

While Fig.~\ref{fig:JCcomp} represents the fidelity error for the full system, this error can be completely explained by the small qubit excited-state population in the Jaynes-Cummings steady-state, shown in Fig.~\ref{fig:EPop}. As previously mentioned, this small excited-state population is due to the weak quantum heating, though the dominant Purcell decay prevents full population inversion. The fidelity of the reduced qubit-states is given by
\begin{align}
	&\nonumber{\rm Qubit\ Fidelity} = {\rm Tr}\left(\sqrt{\sqrt{\rho^Q_{D}}\rho^Q_{JC}\sqrt{\rho^Q_{D}}}\right) \\ \nonumber&= {\rm Tr}\left(\sqrt{\ketbra{g}{g}\left(P_E\ketbra{e}{e} + (1-P_E)\ketbra{g}{g}\right)\ketbra{g}{g}}\right) \\ &= \sqrt{1-P_E},
\end{align}
and this also shown in Fig.~\ref{fig:JCcomp}. As can be seen, the qubit-state error almost exactly estimates the full-state error, with the overestimate at small $\la$ due to numerical error. Therefore, we can conclude that the majority of error in the full state is due to the qubit-state error, and as such the cavity state is not appreciably affected by the higher-order nonlinearities in the Jaynes-Cummings interaction. Finally, we note that the effect of these higher-order terms has previously been investigated for injected squeezing \cite{Elliott:2015aa}.
%%%%%
%%%

\section{Effect of System Losses}
\label{sec:ApLoss}

In this section we examine the effects of inefficiencies in transmission of the measurement signal, which we refer to as external loss, and internal loss from the cavities in the ISTMS setup. We find that both have qualitatively similar effects, and that the ISTMS setups is resilient to moderate amounts of external or internal loss.

\subsection{External Loss}

External loss of the signal is modeled by a beam splitter transformation on the output field $\hat{X}_{-, {\rm out}}(t)$, such that
\begin{align}
\hat{X'}_{-, {\rm out}}(t) = \sqrt{1-\eta}\hat{X}_{-, {\rm out}}(t) + \sqrt{\eta}\hat{z}_{-}(t),
\end{align}
where $\eta$ is the loss coefficient of the beam splitter, and $\hat{z}_{-}(t)$ is vacuum noise entering the signal from the other port of the beam splitter. Using this transformation, it can be shown that the signal and output noise become
\begin{align}
&\mathcal{M}'_S(\tau) = \sqrt{1-\eta}\mathcal{M}_S(\tau), \\
&\mathcal{M}'_N(\tau) = (1-\eta)\mathcal{M}_N(\tau) + \eta\kk\tau,
\end{align}
with $\mathcal{M}_S(\tau)$ and $\mathcal{M}_N(\tau)$ given by their expressions in \ref{sec:inout}. The SNR is now given by
\begin{align}
{\rm SNR}^2_{\rm ext} = \frac{(1-\eta)\left(\mathcal{M}_S(\tau)\right)^2}{2\left[(1-\eta)\mathcal{M}_N(\tau)+ \eta\kk\tau\right]}.
\end{align}
In the long-time limit, and for $\la = \kk/2 - \chi$ this becomes
\begin{align}
{\rm SNR}^2_{\rm ext}  \rightarrow \frac{4\bar{n}_0(1-\eta)}{1 + \eta\left[\frac{1}{2}\left(\frac{\kk}{\chi}\right)^2 - \frac{\kk}{\chi}\right]}\kk\tau.
\end{align}
Therefore, for the SNR to be independent of $\chi$ in the weak coupling limit we require that
\begin{align}
\frac{\chi}{\kk} > \sqrt{\frac{\eta}{2}},
\end{align}
which restricts the amount of external loss the ISTMS setup can tolerate and still change the fundamental scaling of the SNR in the weak coupling limit.

\begin{figure}[h!]
\includegraphics[width=0.8\columnwidth]{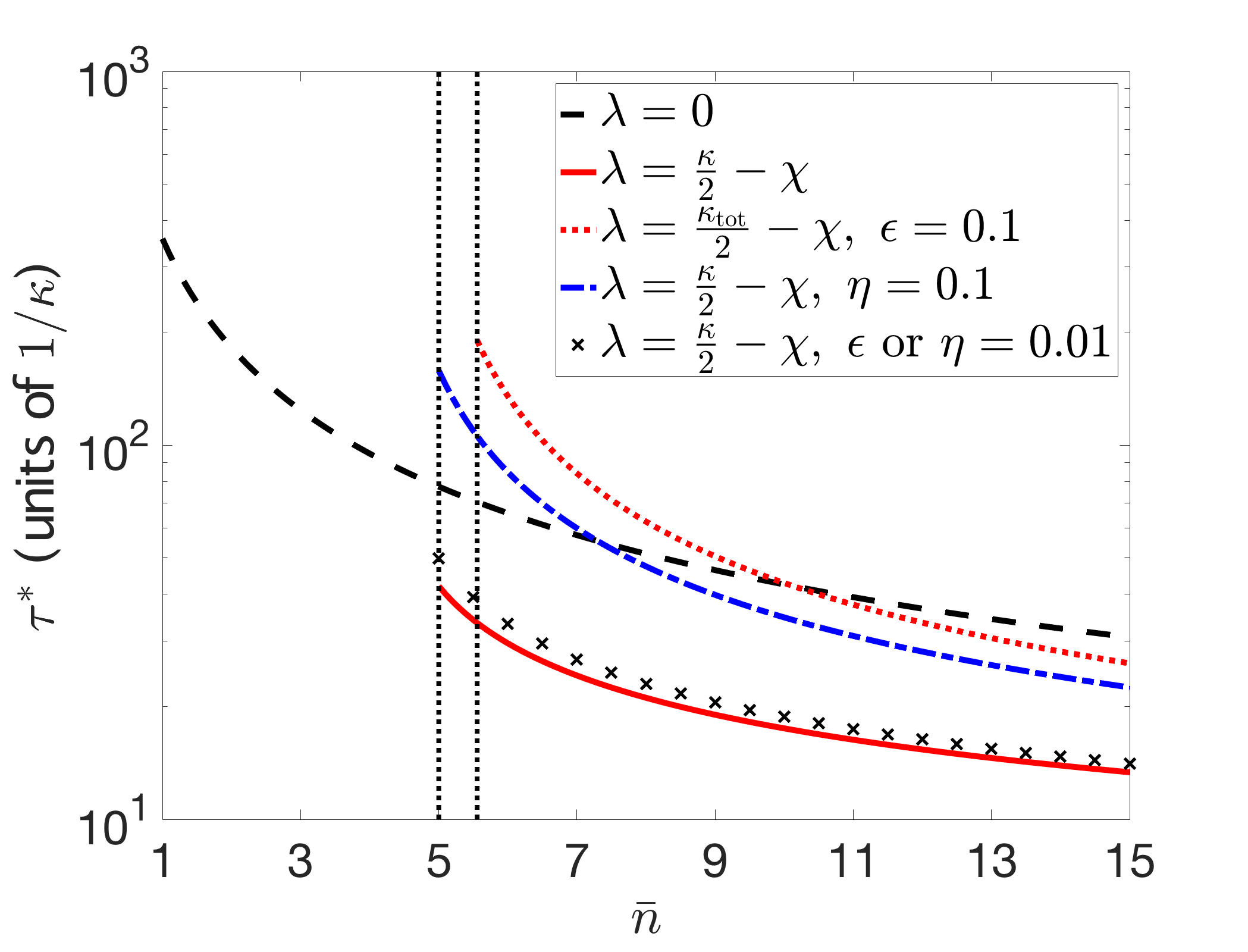}
\caption{Weak coupling ($\chi = \kk/20$) measurement time for various amounts of external ($\eta$) and internal ($\eps$) loss, as a function of total intracavity photon number $\bar{n}$. The output decay rate is $\kk$ for all plots (i.e. $\kk_{\rm tot} = \kk/(1-\eps)$ for the internal loss plots). The dashed vertical lines indicate one plus the number of squeezed photons, with $\bar{n}_{\rm sqz} = 4$ for $\eps =0,0.01$, and $\bar{n}_{\rm sqz} \approx 4.6$ for $\eps =0.1$ (the number of squeezed photons is independent of $\eta$).  $\eta = \eps = 0$ when not otherwise indicated.}
\label{fig:Loss}
\end{figure}

The short time physics is also modified by external loss, and Fig.~\ref{fig:Loss} shows the measurement time for different amounts of external loss. Here, the intuition from the long-time asymptotic limit also holds true, and for sufficiently small loss (in this case $1\%$) the measurement time is almost as short as for no loss. Even for $10\%$ loss the increase in measurement time is less than an order of magnitude over that for no loss.

\subsection{Internal Loss}

For internal loss we introduce the additional input-output channel $\hat{x}_{-,{\rm in/out}}(t)$ with decay rate $\kk_{\rm int}$, which is a fraction $\eps$ of the total decay rate $\kk_{\rm tot} = \kk + \kk_{\rm int}$, such that $\kk_{\rm int} = \eps\left(\kk + \kk_{\rm int}\right)$. The total input state is now described by the quadrature
\begin{align}
\hat{\mathcal{X}}_{-,{\rm in}}(t) = \sqrt{1-\eps}\hat{X}_{-,{\rm in}}(t) + \sqrt{\eps}\hat{x}_{-,{\rm in}}(t).
\end{align}
The input-output relation for the signal output port remains the same
\begin{align}
\nonumber\hat{X}_{-,{\rm out}}(t) &= \hat{X}_{-,{\rm in}}(t) + \sqrt{\kk}\hat{X}_{-}(t) \\&= \hat{X}_{-,{\rm in}}(t) + \sqrt{(1-\eps)\kk_{\rm tot}}\hat{X}_{-}(t),
\end{align}
however, the field inside the cavities, $\hat{X}_{-}(t)$, now depends on $\hat{\mathcal{X}}_{-,{\rm in}}(t)$, not just on $\hat{X}_{-,{\rm in}}(t)$.

Following the same procedure as in \ref{sec:inout}, we calculate the measurement signal and output noise, and we find that the SNR is given by the expression
\begin{align}
{\rm SNR}^2_{\rm int} = \frac{(1-\eps)^2\left(\mathcal{M}_S(\tau)\right)^2}{2\left[(1-\eps)\mathcal{M}_N(\tau)+ \eps\kk_{\rm tot}\tau\right]},
\end{align}
with $\mathcal{M}_S(\tau)$ and $\mathcal{M}_N(\tau)$ as before, but with $\kk$ replaced by $\kk_{\rm tot}$. This is qualitatively similar to the SNR for external loss, and we find that for
\begin{align}
\frac{\chi}{\kk_{\rm tot}} > \sqrt{\frac{\eps}{2}},
\end{align}
the SNR is independent of $\chi$ in the long-time limit for weak coupling and $\la = \kk_{\rm tot}/2-\chi$. For $\eps \ll 1$, $\kk_{\rm tot} \approx \kk$ and this condition is approximately $\chi/\kk > \sqrt{\eps/2}$, which is the same condition as for external loss.

For internal loss the short time physics also follows the behavior predicted by the asymptotic long-time results. As can be seen in Fig.~\ref{fig:Loss}, $10\%$ internal loss is slightly worse than $10\%$ external loss, but at $1\%$ loss the two situations are not discernibly different on the scale of Fig.~\ref{fig:Loss} (and hence they are plotted as one curve).

\section{Qubit Back-Action and Purcell Decay}
\label{sec:ApBA}

\subsection{Density of States}

To obtain the photonic density of states for the left and right cavities, we consider the system evolution in the absence of the qubit in the left/right basis
\begin{align}
&\dot{\hat{a}}_R =  -iJ\hat{\sigma}_z\hat{a}_L - \kk\hat{a}_R - \sqrt{2\kk}\hat{b}_{R,{\rm in}}(t), \\
&\dot{\hat{a}}_L = -i\chi\hat{\sigma}_z\hat{a}_R.
\end{align}
where we have set $\la = 0$ for simplicity, as it does not affect the results for Purcell decay, and we are in a frame such that $\ww_c = \ww_q = 0$. In frequency space, the solution to these equations is given by
\begin{align}
  \left(\begin{array}{c}
  \hat{a}_R(\ww)\\
  \hat{a}_L(\ww)
\end{array}\right) = -\sqrt{2\kk}{\bm \chi}_S(\ww) \left(\begin{array}{c}
  \hat{b}_{R,{\rm in}}(\ww)\\
  0
  \end{array}\right)
\end{align}
where
\begin{align}
    {\bm \chi}_S(\ww) = \frac{1}{i\ww(i\ww + \kk) + J^2}\left(\begin{array}{cc}
      i\ww & -iJ \\
      -iJ & i\ww + \kk
      \end{array}\right)
\end{align}
is the susceptibility matrix for the cavities. The right and left cavity density of states are the real part of the diagonal elements of this matrix, and are given by
\begin{align}
&\rho[\ww]_{\rm R} = 2{\rm Re}\left(\left[{\bm \chi}_S(\ww)\right]_{RR}\right) = \frac{2\kk\ww^2}{\kk^2\ww^2 + \left(J^2-\ww^2\right)^2}, \\
&\rho[\ww]_{\rm L} = 2{\rm Re}\left(\left[{\bm \chi}_S(\ww)\right]_{LL}\right) = \frac{2\kk J^2}{\kk^2\ww^2 + \left(J^2-\ww^2\right)^2}.
\end{align}
Note that this definition is equivalent to defining the density of states in terms of the imaginary part of the retarded cavity Green's function. The density of states is plotted in the main text, and as can be seen from the above equations, the density of states for the right resonator is zero for $\ww = 0$. In the lab frame this corresponds to $\ww = \ww_c = \ww_q$, which explains the suppression of Purcell decay seen for the ISTMS setup.

\subsection{Left Cavity Internal Loss}

We consider internal loss in the left cavity via the system-bath Hamiltonian $\hat{H}'_{SB} = \sqrt{2\kappa'} \hat{B}^{'\dagger} \hat{a}_L + h.c. $, where $\hat{B}^{'\dagger}$ creates an excitation in the left cavity's electromagnetic bath, and $\kk' \ll \kk$ is the internal decay rate. Using the expressions for the eigenmodes in the dispersive regime from the main text, we see that the left cavity transforms as
\begin{align}
\hat{\tilde{a}}_L  \simeq \frac{\hat{a}_E - \hat{a}_O}{\sqrt{2}} - \frac{g}{J}\hat{\sigma}^-,
\end{align}
In this case the two paths for the input noise from the left cavity interfere constructively, and the $\hat{\sigma}^-$ component remains. Therefore, Purcell decay can occur through this channel, with rate $\yy'_{PD} = \kk'\left(g/J\right)^2$. However, since $\kk' \ll \kk$ the Purcell decay rate from the left cavity is very small, and overall, the qubit lifetime has been enhanced by the large factor $\kk/\kk'$ for the ISTMS setup.

\section*{Bibliography}

\bibliographystyle{unsrt}
\bibliography{ISTMSBib}

\end{document}